\documentclass[aps,prl,showpacs,twocolumn,amsmath,amssymb]{revtex4-1}
\usepackage{graphicx,bbm}

\usepackage{amsfonts}
\usepackage{dsfont}
\usepackage{subfigure}
\usepackage{hyperref}

\newcommand{\tr}{\text{tr}}

\def\ket#1{\mathinner{|{#1}\rangle}}

\newcommand{\un}[1]{\underline{#1}}

\newcommand{\ii}{ {\rm i} }
\newcommand{\dd}{ {\rm d} }
\newcommand{\ZZ}{\mathbb{Z}}

\newcommand{\CC}{\mathbb{C}}

\newcommand{\LL}{{\hat {\cal L}}}
\newcommand{\UU}{{\hat{\cal U}}}

\newcommand{\mm}[1]{{\mathbf{#1}}}

\def\one{\mathbbm{1}}

\begin{document}

\title{Non-equilibrium phase transition in a periodically driven XY spin chain}
\author{Toma\v{z} Prosen$^{1,2}$ and Enej Ilievski$^1$}
\affiliation{$^1$Department of Physics, FMF, University of Ljubljana, 1000 Ljubljana, Slovenia}
\affiliation{$^2$ Institute of Physics and Astronomy, University of Potsdam, 14476 Potsdam, Germany}

\date{\today}

\begin{abstract}
We present a general formulation of Floquet states of periodically time-dependent open Markovian quasi-free fermionic many-body systems in terms of a
discrete Lyapunov equation. Illustrating the technique, we  analyze periodically kicked XY spin 1/2 chain which is coupled to a pair of Lindblad reservoirs at its ends. 
A complex phase diagram is reported with re-entrant phases of long range and exponentially decaying spin-spin correlations as some of the system's parameters are varied.
The structure of phase diagram is reproduced in terms of counting non-trivial stationary points of Floquet quasi-particle dispersion relation.
\end{abstract}


\maketitle

{\em Introduction.} Understanding and controlling dynamics of many-body quantum systems when they are open to the environment and driven far from equilibrium is an exciting and important topic of current research in theoretical \cite{reviewT,zoller} and experimental quantum physics \cite{reviewE}.
In particular, since it has been recently realized that certain emergent phenomena, such as quantum phase transitions and long range order -- previously known only in equilibrium zero-temperature quantum states \cite{sachdev} -- can appear also in far from equilibrium steady states of quantum Liouville evolution \cite{zoller,nQPT,pz10,marko}. In investigating dynamical and critical many-body phenomena, quasi-free (quadratic) quantum systems play an important role as they are amenable to analytical treatment (see e.g. \cite{amico}), so many effects can be analyzed exactly or in great detail. For example,
quantum phase transitions in non-equilibrium steady states have been observed either in quasi-free \cite{nQPT,3QRedfield}, or strongly interacting \cite{pz10}, or even dissipative \cite{marko,jens} quantum systems in one dimension. 

At least two distinct types of quantum phase transitions with emergence of long range order have been identified: (i) transitions which result from competition between unitary and dissipative parts of the Liouvillian dynamics \cite{zoller,marko}, and (ii) transitions which occur in boundary driven systems \cite{nQPT,pz10} (without any dissipation in the bulk) as a result some qualitative change in the properties of the unitary dynamics (i.e. of the Hamiltonian), say spontaneous symmetry breaking or bifurcations in quasi-particle dispersion relation \cite{nQPT}. Nevertheless, it has not yet been investigated whether such a structure of long range order can persist in the presence of external, periodic driving. Open and dissipative quantum system periodically driven by intense laser fields have been studied by means of the so called Floquet theory \cite{hanggi}, however most of the previous studies have dealt with essentially single-particle problems. From dynamical system's perspective, it can be argued, that periodically driven or kicked quantum chains are probably the simplest and cleanest examples of quantum many-body ergodic dynamics \cite{tp}. 

Furthermore, periodic time dependence offers new ways of encoding topological information (see e.g. \cite{kitegawa}).
Periodically driven many-body systems have thus been proposed as  candidate systems for studying topological order and detecting Majorana fermions in one dimension \cite{kitegawa,MF}. 
On the other hand, quantum spin chains with particular specific Hamiltonians have been proposed as efficient ways for quantum information transfer \cite{bose}. 
Engineering Hamiltonians and Liouvillians with particular properties in terms of a small set of primitive local operations (realized by laser pulses) is one of
the key problems in designing quantum computation, so it would be highly desirable to have simple and powerful techniques for theoretical treatment of time-dependent open many-body systems, not necessarily time-periodic.

In this Letter, we address the question of exact treatment of periodically driven quasi-free open many-body dynamics, which can be described either by time-dependent Lindblad equation, or equivalently, by discrete dynamical semigroups. We apply the idea of canonical quantization in the Fock space of density operators \cite{original,kosov} and solve for the covariance matrix of the (asymptotic) stationary 
Floquet state in terms of the discrete Lyapunov equation. This yields an efficient general setup applicable to such class of problems in the language of classical control theory \cite{zhou}. Our method, which can be useful also in the case of non-periodic time-dependence, is then applied to treat an open Heisenberg XY spin 1/2 chain which is periodically kicked with a transverse magnetic field. We find an appearance of two distinct phases of spin-spin correlations by changing the (bulk) system's parameters, namely the phase of exponentially decaying correlations and the phase of long range magnetic correlations. The phase diagram of the model is rather complex, but can be reproduced in terms of bifurcations of the Floquet quasi-energy quasi-particle dispersion relation.

{\em Stationary Floquet state.}
We consider a finite system of $n$ (Majorana) fermions, which are described by $2n$ anticommuting Hermitian operators $w_j$, $j=1,\ldots,2n$, $\{w_j,w_k\} = 2\delta_{j,k}$, which may be constructed in terms of Pauli spin 1/2 operators $\sigma_j^{\rm x,y,z}$ via Jordan-Wigner transformation
$w_{2m-1} = \sigma^{\rm x}_m \prod_{m' < m} \sigma^{\rm z}_{m'}, w_{2m} = \sigma^{\rm y}_m \prod_{m' < m} \sigma^{\rm z}_{m'}.$
We shall be interested in a solution of a general time-dependent Markovian master equation for the system's density operator $\rho(t)$,
\begin{equation}
\frac{\dd}{\dd t} \rho(t) = \hat{\cal L}(t)\rho(t),
\quad
\hat{\cal L}(t)\rho\equiv 
-\ii [H(t),\rho] + \hat{\cal D}(t) \rho,
\label{eq:lindblad}
\end{equation}
where the dissipator may either be of a Lindblad form $\hat{\cal D}\rho = \sum_\mu (2L_\mu \rho L_\mu^\dagger - \{L^\dagger_\mu L_\mu,\rho\})$, given in terms of (possibly time dependent) Lindblad operators $L_\mu(t)$, or even of more general, Redfield form 
(see Ref.~\cite{3QRedfield} for the formulation in compatible notation). We focus on quasi-free dynamics where the Hamiltonian is given in terms of a quadratic form $H = \sum_{j,k} w_j H_{j,k} w_k \equiv \un{w}\cdot\mm{H}\un{w}$ with antisymmetric imaginary matrix $\mm{H}$ and
linear Lindblad operators $L_\mu = \sum_{j} l_{\mu,j} w_j \equiv \un{l}_\mu \cdot \un{w}$, $\un{l}_\mu \in \CC^{2n}$. Here and below, hatted symbols (such as the Liouvillian $\LL$) designate
super-operators over $4^n$ dimensional operator (Liouville) space, and bold-roman symbols (such as $\mm{H}$) designate $2n\times 2n$ or $4n\times 4n$ matrices.

Recently, a general approach to explicit analysis of dynamical properties of Markovian master equations of open many-body systems has been proposed \cite{original,3QRedfield,spectral}, which is based on quantization in the Fock space of (density) operators, sometimes referred to as the `third quantization'. In the context of quasi-free systems, if one is only interested in pair-correlations, dynamics of the covariance matrix 
$C_{j,k}(t) := \tr[w_j w_k \rho(t)] - \delta_{j,k}$ can be derived, either directly from (\ref{eq:lindblad}) by observing {\em canonical anticommutation relations} (CAR) among $w_j$, 
or by means of fermionic super-operators over the Liouville space \cite{original,3QRedfield}
\begin{equation}
\dot{\textbf{C}}(t) =-\textbf{X}(t)\textbf{C}(t)-\textbf{C}(t)\textbf{X}^{T}(t)+ \ii \mm{Y}(t).
\label{eq:tdepC}
\end{equation}
$\textbf{X}(t) = 4(\ii\textbf{H}(t)+\textbf{M}_{\rm r}(t))$ and $\mm{Y}(t) = 4(\mm{M}_{\ii}(t)-\mm{M}^T_{\ii}(t))$ are real $2n\times 2n$ time-dependent matrices,
where $\mm{M}\equiv \mm{M}_{\rm r} + \ii \mm{M}_{\ii}$ is a bath matrix, given as $\mm{M} = \sum_\mu \un{l}_\mu \otimes \bar{\un{l}}_\mu$ for the Lindblad model, whereas the expressions for the Redfield model can be found in Refs.~\cite{3QRedfield,3QExplicit}. In the time-independent case, one is interested in the steady state solution of (\ref{eq:tdepC}) $\dot{\mm{C}}=0$, completely determined via the solution of the continuous Lyapunov equation $\textbf{X}\textbf{C}+\textbf{C}\textbf{X}^{T}=\ii\mm{Y}$.
For a general, time-dependent case, the general solution of (\ref{eq:tdepC}) for covariances can be sought for in terms of an ansatz
\begin{equation}
\textbf{C}(t)=
\textbf{Q}(t)\textbf{C}(0)\textbf{Q}^{T}(t) - \ii\textbf{P}(t)\textbf{Q}^{T}(t),
\label{eq:evol}
\end{equation}
which results in two simpler equations, for $2n\times 2n$ real matrix functions $\mm{P}(t),\mm{Q}(t)$
\begin{eqnarray}
\dot{\mm{Q}}(t) &=& -\mm{X}(t)\mm{Q}(t), \qquad\qquad\qquad\quad \mm{Q}(0) = \mm{1}, \label{eq:PQdif} \\
\dot{\mm{P}}(t) &=& -\mm{X}(t)\mm{P}(t) - \mm{Y}(t)\mm{Q}^{-T}(t), \;\; \mm{P}(0) = \mm{0}, \nonumber
\end{eqnarray}
with explicit solutions
\begin{eqnarray}
\mm{Q}(t) &=& \hat{\cal T}\exp\left(-\int_{0}^{t}\!\!\dd t'\, \mm{X}(t')\right), \label{eq:PQ}\\
\mm{P}(t) &=& -\mm{Q}(t)\int_0^t \!\dd t' \mm{Q}^{-1}(t')\mm{Y}(t')\mm{Q}^{-T}(t'). \nonumber
\end{eqnarray}

In this Letter we are interested in a time-periodic Markovian master equation (\ref{eq:lindblad}), with $\LL(t+\tau) = \LL(t)$, and looking for the stationary state $\rho_{\rm F}$, satisfying
\begin{equation}
\UU(\tau,0) \rho_{\rm F} = \rho_{\rm F}, \quad \UU(t'',t') = \hat{\cal T}\exp\left(\int_{t'}^{t''}\!\!\!\!\dd t\, \LL(t)\right),
\end{equation}
which we will refer to as {\em stationary Floquet state} (SFS).
The covariance matrix in SFS, $\mm{C}_{\rm F}$, can be given in terms of a solution of the {\em discrete Lyapunov equation} \cite{zhou}
\begin{equation}
\mm{Q}(\tau) \mm{C}_{\rm F} - \mm{C}_{\rm F}\mm{Q}^{-T}(\tau) = \ii \mm{P}(\tau),
\label{eq:disLyap}
\end{equation}
which results from (\ref{eq:evol}) after plugging $\mm{C}(0)=\mm{C}(\tau)=\mm{C}_{\rm F}$.
It could be of some interest also to understand the structure of the Floquet-Liouville spectrum of the many-body super-operator $\UU(\tau,0)$. We show in appendix to this Letter \cite{appendix} that a complete spectrum of 
$\UU(\tau,0)$ can be written in terms of eigenvalues of the matrix $\mm{Q}(\tau)$. For example, the eigenvalue $\lambda_1$ of $2n\times 2n$ matrix $\mm{Q}(\tau)$ of maximal modulus  (note that all eigenvalues of $\mm{Q}$ have modulus less than $1$), is also a maximal modulus eigenvalue of $4^n \times 4^n$ matrix $\UU(\tau,0)$, and determines the time-scale $t^* = \tau/\log|1/\lambda_1|$ of relaxation to SFS.

\begin{figure}
\centering	
\includegraphics[width=\columnwidth]{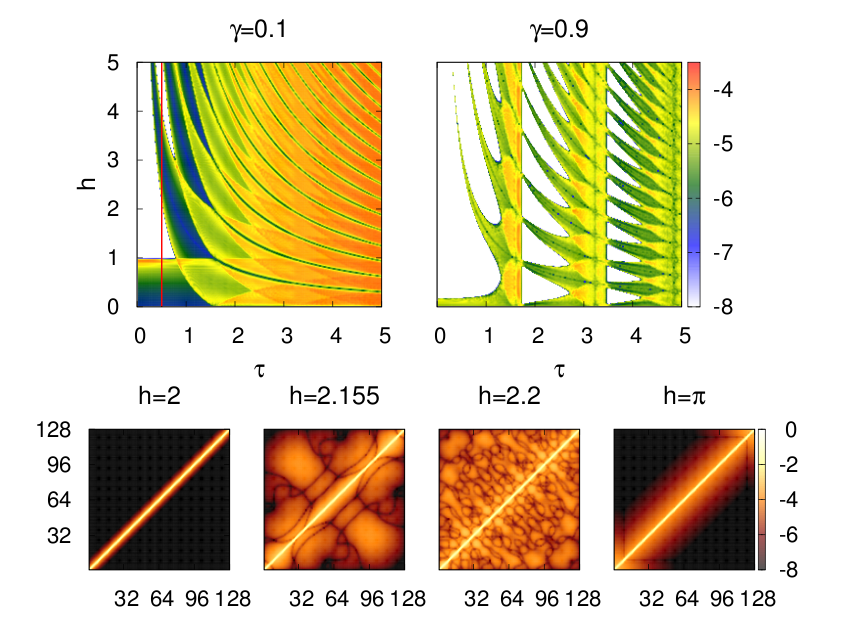}
\caption{Residual correlations $C_{\text{res}}(\tau,h)$ (plotted in a log-scale, $n=64$, with 
coupling constants $\Gamma_{1}^{L}=\Gamma_{1}^{R}=0.5$, $\Gamma_{2}^{L}=0.3$ and 
$\Gamma_{2}^{R}=0.1$) in SFS of kicked Heisenberg XY spin  chain for anisotropy
$\gamma=0.1$ (top left) and $\gamma=0.9$ (top right). Correlations that fall below
$10^{-8}$ threshold belong to white colored region and correspond to a phase
with exponentially decaying correlations. At higher anisotropy LRMC phase ceases to
be a simply connected region. Below we show some typical spin-spin correlation matrices chosen
along the red line on $\gamma=0.1$ phase diagram ($n=128$). Leftmost panel clearly shows exponential fall-off from the diagonal, the middle two belong to LRMC phase near
the phase boundary, and the rightmost to a kind of \textit{antiresonance} which occurs for $h = p \pi/(2\tau), p \in\ZZ$ where the kick has no effect.}
\label{fig:figure1}
\end{figure}

{\em Kicked open XY spin chain.} Let us now consider a special case of periodically kicked systems with time dependent structure matrix 
$\mm{X}(t) = \mm{X}^0 + \delta_\tau(t) \mm{X}^1$ and time independent $\mm{Y} =  8\mm{M}_\ii$, where $\delta_\tau(t) \equiv \tau\sum_{m\in\ZZ} \delta(t-m\tau)$ is a periodic Dirac 
function with period $\tau$. Starting the one-period time interval just after the kick, we find explicit solutions of Eqs. (\ref{eq:PQdif})
\begin{eqnarray}
\mm{Q}&=&\exp(-\tau\mm{X}^1)\exp(-\tau\mm{X}^0), \label{eq:PQkicked}\\
\mm{P}&=&\mm{Q}(\exp(\tau\mm{X}^0)\mm{Z} \exp(\tau \mm{X}^0)^T - \mm{Z}), \nonumber
\end{eqnarray}
where $2n\times 2n$ real antisymmetric matrix $\mm{Z}$ is a solution of the {\em continuous Lyapunov equation}
\begin{equation}
\mm{X}^0 \mm{Z} + \mm{Z} (\mm{X}^0)^T = \mm{Y}.
\label{eq:conLyap}
\end{equation}
Note that $-\ii\mm{Z}$ is the covariance matrix for the time-independent case \cite{spectral}.
In the following we apply our method to study the kicked XY Heisenberg spin 1/2 chain
\begin{equation*}
H(t) = \sum_{j=1}^{n-1} \left(\!\frac{1\!+\!\gamma}{2} 
\sigma^{\rm x}_j \sigma^{\rm x}_{j+1} + \frac{1\!-\!\gamma}{2} \sigma^{\rm y}_j \sigma^{\rm y}_{j+1}\!\right)
+ h \delta_\tau(t) \sum_{j=1}^n \sigma^{\rm z}_j,
\end{equation*} 
coupled to a pair of Lindblad baths at its ends, with four
 Lindblad operators $L_{1,2} = \sqrt{\Gamma_{1,2}^{\rm L}} \sigma^{\pm}_1$,
$L_{3,4} = \sqrt{\Gamma_{1,2}^{\rm R}} \sigma^{\pm}_n$. Hence, the model is quadratic in Majorana fermions $w_j$~\cite{original},
and the structure matrices can be written in $2\times 2$ block-matrix form as
\begin{eqnarray}
\mm{X}^0_{j,k} &=&(\gamma \sigma^{\rm x} -\ii \sigma^{\rm y}) \delta_{j,k-1} - (\gamma\sigma^{\rm  x} + \ii \sigma^{\rm y})\delta_{j,k+1} \nonumber \\
&+& \Gamma^{\rm L}_+ \one_2 \delta_{j,1}\delta_{k,1} + \Gamma^{\rm R}_+ \one_2  \delta_{j,n}\delta_{k,n},\nonumber \\
\mm{X}^1_{j,k} &=& 2\ii h \sigma^{\rm y} \delta_{j,k} ,\\ 
\mm{Y}_{j,k} &=& 2\ii \Gamma^{\rm L}_- \sigma^{\rm y} \delta_{j,1}\delta_{k,1} +2\ii \Gamma^{\rm R}_- \sigma^{\rm y}\delta_{j,n}\delta_{k,n} \nonumber
\end{eqnarray}
 where $\Gamma^{\rm L,R}_{\pm} \equiv \Gamma^{\rm L,R}_2 \pm  \Gamma^{\rm L,R}_1$ and
$j,k=1,\ldots,n.$
Using standard linear algebra routines we have solved numerically the Eq. (\ref{eq:disLyap}) with (\ref{eq:PQkicked},\ref{eq:conLyap}) for different values of the system's parameters and 
different chain sizes up to $n\sim 10^3$.
In particular we have focused on the existence of long-range order in SFS of the model, so we have defined an order parameter -- residual correlator $C_{\rm res}=(\sum_{j,k}^{|j-k|\ge n/2} |C_{j,k}|)/(\sum_{j,k}^{|j-k|\ge n/2} 1)$.
In accordance with the time-independent open XY model \cite{nQPT} we have discovered regions in parameter space $(\gamma,\tau,h)$ where $C_{\rm res}$ decays exponentially with $n$, and regions of parameters where we have correlations over large distances and $C_{\rm res}\propto 1/n$ (the phase of {\em long range magnetic correlations} (LRMC)). However, in contrast with the time-independent case, 
we find here a very rich phase diagram (see Fig.\ref{fig:figure1}) with re-entrant LRMC phase.

\begin{figure}
\centering	
\includegraphics[width=\columnwidth]{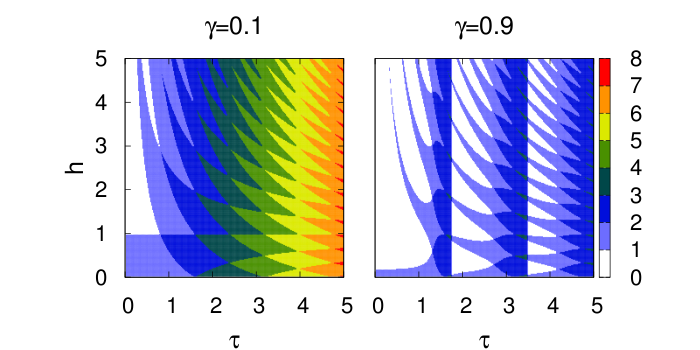}
\caption{Half the total number of non-trivial stationary points (as their number is always even) of quasiparticle dispersion $\theta(\kappa)$ for kicked
XY chain (same parameters as in Fig.~\ref{fig:figure1}). The structure evidently
coincides with the phase diagram of residual correlations $C_{\text{res}}(\tau,h)$.}
\label{fig:figure2}
\end{figure}

We find that qualitative properties of the phase diagram are completely independent of the bath parameters $\Gamma^{\rm L,R}_{1,2}$,
so we expect that theoretical explanation rests only on the bulk properties. In analogy to the time-independent case, where the phases have been explained based on
quasi-particle dispersion relation of the infinite XY chain~\cite{nQPT}, we study dispersion relation for the Floquet modes $\mm{Q} \un{u} = e^{\ii \theta} \un{u}$ in the absence of the baths ($\mm{P}=\mm{0}$).
Due to translational invariance of the infinite chain, the Floquet modes 
can be written in terms of the Bloch waves $\un{u}_j = \un{a}(\kappa) \exp(\ii j \kappa)$, 
namely $\mm{T}(\kappa)\un{a}(\kappa)= e^{\ii\theta(\kappa)}\un{a}(\kappa)$,
where 
\begin{equation}
\mm{T}(\kappa) = \exp[-\tau \mm{X}^1_{0,0}]\exp[-\tau(\mm{X}^0_{1,0}e^{\ii \kappa}+\mm{X}^0_{0,1}e^{-\ii \kappa})]
\end{equation}
is a SU(2) matrix, with two quasi-energy bands 
$\theta_{1,2}(\kappa)=\pm \arccos\!\left[\cos(2\tau h)\!\cos(2\tau\epsilon(\kappa)) + \sin(2\tau h)\!\sin(2\tau \epsilon(\kappa))\frac{\cos\kappa}{\epsilon(\kappa)}\right]$
and $\epsilon(\kappa) = \sqrt{\cos^2\kappa + \gamma^2 \sin^2\kappa}$ is the quasi-particle energy for the un-kicked XY chain, $\kappa\in[-\pi,\pi)$.
As it has been argued in Ref.~\cite{nQPT}, LRMC phase is signaled by the appearance of a nontrivial stationary point $\kappa^*$ in the quasi-particle dispersion relation, i.e. $\dd \theta(\kappa^*)/\dd\kappa = 0$ for $\kappa^* \not\in \{0,\pi\}$.
In fact, near the phase boundary the nontrivial stationary points are emerging, say at $\kappa=0$, so the critical parameters can be obtained via Landau
scenario by expanding the quasi-energy  
 $\theta(\kappa) =  \theta_0+a(\gamma,\tau,h)\kappa^{2}+b(\gamma,\tau,h)\kappa^{4} + {\cal O}(\kappa^6)$. 
 Then, the nontrivial solution $0\neq \kappa^* =\pm \sqrt{-a(\gamma,\tau,h)/2b(\gamma,\tau,h)}$ exists when $a$ and $b$ have different signs, thus a
 new pair of stationary points emerges when
$a(\gamma,\tau,h)\equiv\tau(1-\gamma^{2})+\frac{1}{2}\gamma^{2}\sin{(2\tau)}\sin{(2\tau h)}/\sin{(2\tau(h-1))}=0$, which defines the critical field $h_{\rm c}$ by solving the equation $a(\gamma,\tau,h_{\rm c}(\gamma,\tau))=0$.
Such curves $h_{\rm c}(\gamma,\tau)$ exactly reproduce the rib structure of the phase diagrams in Figs.~\ref{fig:figure1},\ref{fig:figure2} and 
-- where separating the regions between 0 and 2 nontrivial stationary points -- define the phase boundary.
Furthermore, in Fig.~\ref{fig:figure2} we observe that the phase diagram structure can be nicely reproduced by plotting the number $n_\sharp$ of non-trivial stationary points $\kappa^*$ of
$\theta(\kappa)$, the total number of stationary points being $n_\sharp+2$. 
This has interesting consequences on the temporal growth of block entropy in the operator space \cite{iztok} which is conjectured as $S(t) \sim \frac{1}{3}(n_\sharp+2) \ln t$.

\begin{figure}
\centering	
\includegraphics[width=\columnwidth]{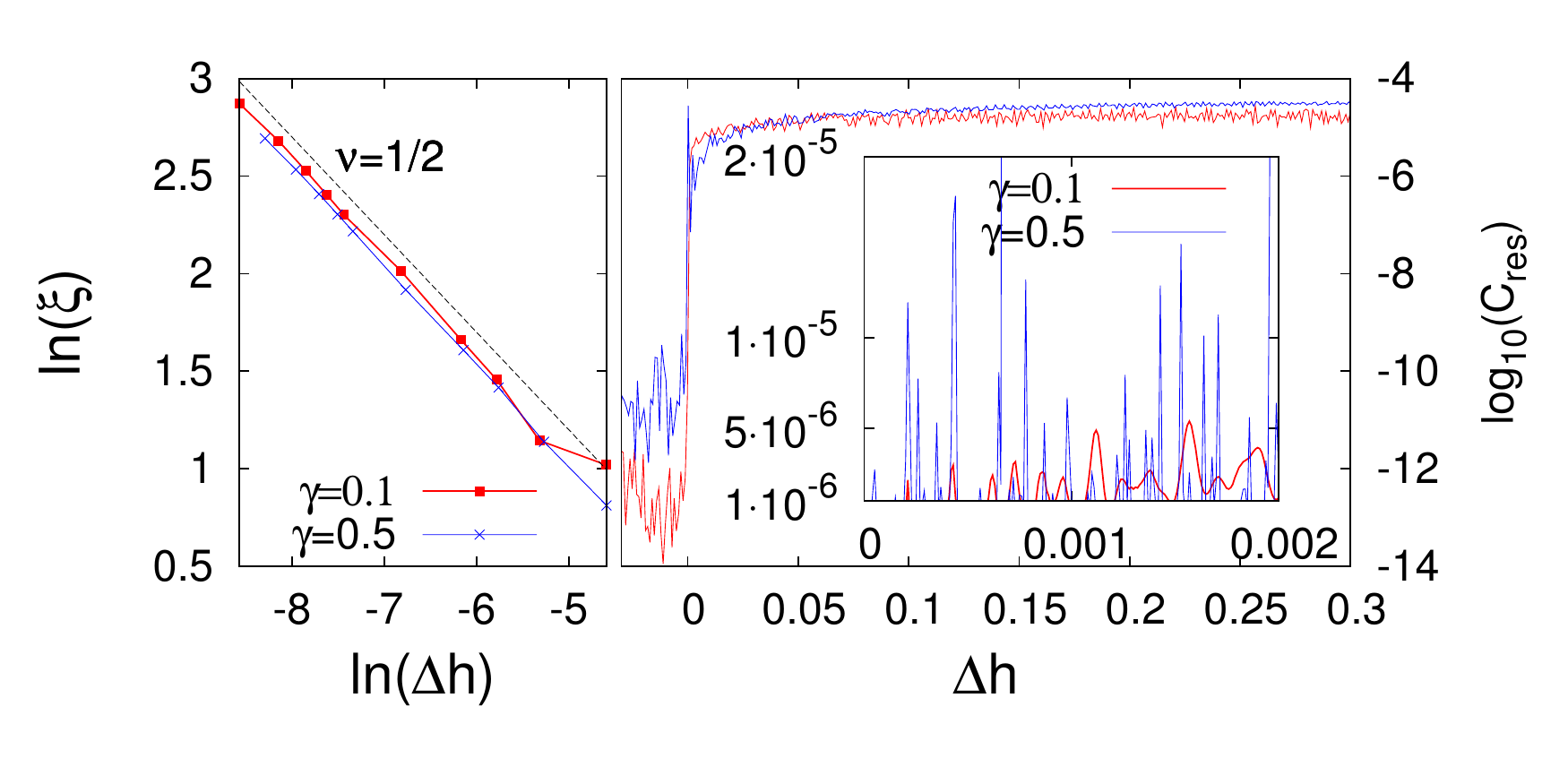}
\caption{Critical behaviour of correlations in non-LRMC phase (left) and LRMC phase (right) for kicked XY chain of size
 $n=480$, and $\tau=0.5$, where $\Delta h=h-h_{\rm c}$. In non-LRMC phase correlations decay as 
$C_{\text{res}}(r)\propto \exp{(-r/\xi)}$. We confirm, for two values of $\gamma=0.1,0.5$ that correlation length $\xi$ diverges with critical 
exponent $\nu=1/2$ when approaching the critical point.
In the right panel we show an abrupt transition of the order parameter $C_{\rm res}$ when entering the LRMC phase. In the inset,  zooming in the region near the phase transition point, we observe appearance of the correlation resonances (explained in the simplified context in Ref.~\cite{3QExplicit}). Couplings $\Gamma^{\rm L,R}_{1,2}$ are reduced by a factor of $10^{-6}$ w.r.t.
Fig.~\ref{fig:figure1}.}
\label{fig:figure3}
\end{figure}

One can define a critical exponent by approaching the phase boundary from the side where
correlations decay exponentially $C_{j,k} \sim \exp (-|j-k|/\xi)$, $\xi \sim |h-h_{\rm c}|^{-\nu}$. Using our heuristic quasi-particle picture,
$\nu$ can be derived from the imaginary stationary point wave-number near the critical point
$(\gamma,\tau,h_{\rm c}(\gamma,\tau))$,
$
\kappa^* \sim \ii |h~-~h_{\rm c}(\gamma,\tau)|^{1/2},
$
giving the critical exponent $\nu=1/2$, which is confirmed by numerical simulations in kicked XY chain in  Fig.~\ref{fig:figure3}.
We have carefully checked numerically another signature of non-equilibrium quantum phase transition, namely the closing of the spectral gap $\Delta(n)=-\ln|\lambda_1|$. We find clearly
-- similarly as in time-independent open XY model~\cite{nQPT,3QRedfield} -- that $\Delta(n) \propto n^{-3}$ away from the phase boundary
$h\neq h_{\rm c}$ whereas $\Delta(n) \propto n^{-5}$ for $h=h_{\rm c}$.

{\em Conclusions.} We presented a general setup for treating periodically driven open quasi-free many-body dynamics in terms of quantization in the Liouville-Fock space \cite{diploma}. It is shown that 
covariance matrix for the stationary Floquet state satisfies a discrete Lyapunov equation, well known in classical theory of differential equations and in control theory \cite{zhou}.
Applying the method to study kicked open XY spin chain, we find a rich phase diagram of re-entrant phases of long-range correlation order and exponentially decaying correlations
upon variation of a generic system's parameter.

We acknowledge support by the grants P1-0044, J1-2208 of Slovenian research agency, as well Bessel research prize of A. v. Humboldt foundation.

\newpage

\section*{Supplementary Material: \\ 
Floquet-Liouvillian spectrum of decay modes.} 

Let us now briefly show how to construct the complete spectrum of decay modes of our quasi-free Floquet-Liouvillian theory. Following Refs.~\cite{original,3QRedfield,spectral} we define $4n$ Hermitian Majorana maps $\un{\hat a}=(\un{\hat a}_1,\un{\hat a}_2)$, as $\hat{a}_{1,j} = (\hat{c}_j +\hat{c}^\dagger_j)/\sqrt{2}$,  
$\hat{a}_{2,j} = \ii(\hat{c}_j -\hat{c}^\dagger_j)/\sqrt{2}$, where canonical fermionic maps are uniquely defined by their action
$\hat{c}_j \ket{P_{\un{\alpha}}} = \delta_{\alpha_j,0} \ket{w_j P_{\un{\alpha}}}$,
$\hat{c}^\dagger_j \ket{P_{\un{\alpha}}} = \delta_{\alpha_j,1} \ket{w_j P_{\un{\alpha}}}$  over the complete Fock basis $P_{\un{\alpha}} = 2^{-n/2}w_1^{\alpha_1}\cdots w_{2n}^{\alpha_{2n}}$ of the
operator space. Referring to Sect. 3.5 of \cite{3QRedfield} we find the Liouvillian propagator 
\begin{equation}
\hat{\cal U}(t,0) = N(t)\exp\left({\textstyle \frac{1}{2}}\un{\hat a}\cdot\left[\ln\mm{U}(t)\right]\un{\hat a}\right),
\label{eq:LFmap}
\end{equation}
where, after a tedious but direct calculation, one finds 
\begin{equation}
\mm{U}(t) = \mm{S}^\dagger 
\begin{pmatrix}
\mm{Q}(t) & \ii \mm{P}(t) \cr
\mm{0} & \mm{Q}^{-T}(t) 
\end{pmatrix} \mm{S},\;\,
\mm{S} \equiv \frac{1}{\sqrt{2}}
\begin{pmatrix}
\mm{1} & -\ii\mm{1}\cr \mm{1} & \ii\mm{1}
\end{pmatrix} ,
\end{equation}
with $\mm{P}(t),\mm{Q}(t)$ given by (\ref{eq:PQ}) and $N(t)=[\det\mm{Q}(t)]^{1/2}$.
Write shortly $\mm{Q}\equiv \mm{Q}(\tau)$, $\mm{P}\equiv\mm{P}(\tau)$, $\mm{U}=\mm{U}(\tau)$, and
assume that $\mm{Q}$ can be diagonalized  \cite{note},
$\mm{Q} = \mm{R}\mm{D}\mm{R}^{-1}$, with
$\mm{D}={\rm diag}\{ \lambda_1,\ldots,\lambda_{2n}\}$.
Then, $\mm{U}$ is diagonalized as
\begin{equation}
\mm{U} = \mm{V}^{-1}(\mm{D}\oplus\mm{D}^{-1})\mm{V},\;
\mm{V} = \begin{pmatrix}
\mm{R}^{-1} & \mm{R}^{-1}\mm{C}_{\rm F}\cr
\mm{0} & \mm{R}^{T}
\end{pmatrix} \mm{S},
\end{equation}
where $\mm{V}^{-1} = \mm{V}^T \mm{J}$, $\mm{J}\equiv \sigma^{\rm x}\otimes \mm{1}_{2n}$.
Now, in exact analogy with the time-independent case, we define normal master mode maps
$(\un{\hat b},\un{\hat b}') = \mm{V}\un{\hat a}$, or 
\begin{equation}
\un{\hat b}=\mm{R}^{-1}(\un{\hat c} + \mm{C}_{\rm F} \un{\hat c}^\dagger),\quad
\un{\hat b}'=\mm{R}^T \un{\hat c}^\dagger ,
\end{equation}
satisfying CAR, $\{\hat{b}_j,\hat{b}_k\}=\{\hat{b}'_j,\hat{b}'_k\}=0$, $\{\hat{b}_j,\hat{b}'_k\}=\delta_{j,k}$
which diagonalize the many-body Liouville-Floquet map
\begin{equation}
\hat{\cal U}(\tau,0) = \exp\left(\sum_{j=1}^{2n} (\ln \lambda_j) \hat{b}'_j \hat{b}_j\right).
\end{equation}
As dynamics should be stable we have $\forall j, |\lambda_j| \le 1$.
The complete set of $4^n$ Floquet modes, $\UU(\tau,0)\ket{\un{\nu}} = \Lambda_{\un{\nu}}\ket{\un{\nu}}$,  is then
$\ket{\un{\nu}} = \prod_{j=1}^{2n} (\hat{b}'_j)^{\nu_j} \ket{\rho_{\rm F}}$, $\nu_j\in\{0,1\}$, with decay rates
$\Lambda_{\un{\nu}} = \prod_j \lambda_j^{\nu_j}.$ Note that the {\em largest modulus eigenvalue} $\lambda_1$ of $\mm{Q}$ 
determines the {\em spectral gap} of $\UU(\tau,0)$, or the relaxation time to $\rho_{\rm F}$ as $t^* = \tau/(-\ln|\lambda_1|)$.

We note that the covariance flow (\ref{eq:tdepC}) describes any, possibly non-Gaussian, time-dependent state. Nevertheless, SFS $\ket{\rho_{\rm F}}$ can be thought of as the right vacuum state of a non-Hermitian quadratic periodically time-dependent field theory, determined by $\hat{b}_j\ket{\rho_{\rm F}}=0$, which implies that $\ket{\rho_{\rm F}}$ {\em is} a Gaussian state, and all higher-order observables can be expressed in terms of covariances $\mm{C}_{\rm F}$ via the Wick theorem.


\begin{thebibliography}{10}
\bibitem{note}
The more general non-diagonalizable case can be treated by means of Jordan decomposition of $\mm{Q}$ in a way analogous to Ref.~\cite{spectral}.
\end{thebibliography}
\end{document}